\documentstyle[epsf,prl,aps]{revtex}
\begin{document}
\twocolumn[%
\hsize\textwidth\columnwidth\hsize\csname@twocolumnfalse\endcsname
\title{%
\hfill{\normalsize%
\raisebox{0pt}[0pt][0pt]%
{\vbox{\hbox{June 1999} \hbox{DPNU-99-21} }}%
}\\
\bf Conformal Phase Transition and 
Fate of the Hidden Local Symmetry\\
in Large $N_f$ QCD}
\author{\bf Masayasu Harada and Koichi Yamawaki}
\address{Department of Physics, Nagoya University,
Nagoya 464-8602, Japan.}
\maketitle

\begin{abstract}
It is observed that the Hidden Local Symmetry (HLS) for the vector
mesons in the ordinary QCD with smaller $N_f$ 
plays the role of the ``Higgsed magnetic gauge symmetry''
for the Seiberg duality in the SUSY QCD. For large $N_f$ where
the conformal phase transition with chiral restoration and
deconfinement is expected to take place, we find that the HLS model
also exhibits the chiral restoration by the loop corrections 
(including the quadratic divergence) in a manner similar to that
in the $CP^{N-1}$ model, provided that the bare HLS Lagrangian
respects the Georgi's vector limit at a certain $N_f$ ($\approx 7$).
\end{abstract}
\vskip1pc]

Increasing attention has been paid to the duality 
in various contexts of modern particle theory. 
Seiberg found the ``electric-magnetic'' 
duality in ${\cal{N}}=1$ Supersymmetric 
(SUSY) QCD with $N_c$ colors and
$N_f$ flavors~\cite{Seiberg}:
For the region $\frac{3}{2} N_c < N_f < 3 N_c$
(``conformal window'') in the SUSY QCD,
there exists a ``magnetic theory'' with the SU($N_f-N_c$) 
gauge symmetry which is dual to the original 
SU($N_c$) theory regarded as the ``electric theory''.
Although the origin of the magnetic gauge symmetry 
(``induced at the composite level'') is not obvious from
the original theory,
both theories in fact have the infrared (IR) fixed point 
with exact conformal symmetry and with the same IR physics.
When $N_f$ decreases, 
the electric theory becomes stronger in IR, while
the magnetic theory gets weaker, with the magnetic
gauge group being reduced through the Higgs mechanism. 
Decreasing $N_f$ further beyond the conformal window, we finally
arrive at $N_f=N_c$ where the magnetic theory is in complete Higgs
phase (reduced to no gauge group), which corresponds to the complete
confinement (and spontaneously broken chiral symmetry) of the electric theory.

Similar conformal window may also exist in the ordinary (non-SUSY)
QCD with massless $N_f$ flavors. 
There actually exists an IR fixed point at two-loop beta function
for large $N_f$ ($< \frac{11}{2} N_c$): 
When $N_f$ increases close to the 
point $\frac{11}{2} N_c$,
the coupling at 
the IR fixed point becomes very small 
so that the deconfinement and the chiral symmetry restoration 
are expected to occur~\cite{BZ}. 
Based on this IR fixed point, it was found~\cite{ATW}
through the modified ladder Schwinger-Dyson (SD) equation
that chiral symmetry restoration 
in fact takes place for $N_f$ 
such that $N_f^{\rm cr} < N_f < \frac{11}{2} N_c$, where
$N_f^{\rm cr}\simeq 4 N_c$($=12$ for $N_c=3$).
In Ref.~\cite{MY}
this chiral restoration at $N_f^{\rm cr}$
was further identified with 
``conformal phase transition'' which was characterized by the 
essential singularity scaling.
Moreover, the lattice simulation indicates that the chiral restoration 
does occur at $N_f^{\rm cr} \approx 7$~\cite{IKKSY}.

Here we recall that, for small $N_f$, 
the vector mesons such as the $\rho$ meson 
can be regarded as the dynamical gauge bosons of Hidden Local
Symmetry (HLS)~\cite{BKUYY,BKY}
in the nonlinear sigma model (Chiral Lagrangian).
The HLS is completely broken 
through the Higgs mechanism as the origin of the 
vector meson mass. 
This gauge symmetry is
induced at the composite level and has nothing to do with the 
fundamental color gauge
symmetry. Instead, the HLS is associated with
the flavor symmetry. 

In this paper we shall find that the Seiberg duality is realized also
in the ordinary (non-SUSY) QCD through the HLS.
This will shed new light on the non-perturbative dynamics of the 
real-life QCD.

We first observe that, for small $N_f$,
the SU($N_f$) HLS is in complete
Higgs phase and yields the same IR physics as the SU($N_c$) QCD in the
confinement/chiral-symmetry-breaking phase, and plays the role of
the ``Higgsed magnetic gauge theory'' 
dual to the ``confined electric gauge theory'' (QCD)
in the spirit of Seiberg duality.

What then happens to the HLS when $N_f$ becomes large so that 
QCD undergoes the conformal phase transition into the conformal
window with deconfinement/chiral restoration?
In order for the duality between QCD and the HLS
be consistently satisfied,
there should be a way 
that the chiral restoration takes place for large $N_f$
also in the HLS theory 
{\it by its own dynamics}.
Actually, it is known
that, in the 
$CP^{N-1}$ nonlinear sigma model 
based on the coset space 
SU($N$)$/$SU($N-1$)$\times$U(1),
the SU($N$) symmetry is restored by the loop effects
and the U($1$) gauge symmetry is dynamically generated accordingly
(See, e.g., Ref.~\cite{BKY}.).
This suggests that
the HLS can provide the chiral restoration by its own dynamics.
In other words, due to the dynamics of the HLS
{\it
the quantum theory is in the symmetric phase even if
the bare theory is written as if it were in the broken phase}.

Now our task is to find a condition for the bare theory of the HLS to
realize chiral restoration for large $N_f$ in the quantum theory.
One clue is the fact~\cite{HY}
that through the renormalization-group equations (RGE's) 
the HLS approaches to the
Georgi's vector limit~\cite{Georgi}
in the idealized 
high energy limit.
The vector limit
is actually the ultraviolet fixed point of the RGE's.

We then propose that 
taking the Georgi's vector limit~\cite{Georgi} in the bare theory 
of the HLS 
at a certain critical value $N_f^{\rm cr}$
is a consistent way to incorporate the conformal
phase transition into the HLS.
In other words,
{\it
the quantum theory provides the chiral restoration
when the bare theory approaches to the vector limit 
as $N_f \rightarrow N_f^{\rm cr}$}.

Let us first describe the HLS model based on the
$G_{\rm global}{\times}H_{\rm local}$ symmetry, where
$G = \mbox{SU($N_f$)}_L \times \mbox{SU($N_f$)}_R$  is the 
global chiral symmetry
and $H = \mbox{SU($N_f$)}_V$ is the HLS.
(The flavor symmetry is the diagonal sum of $G_{\rm global}$
and $H_{\rm local}$.)
The Lagrangian is~\cite{BKUYY,BKY}
\begin{equation}
{\cal L} = F_\pi^2 \, \mbox{tr} 
\left[ \hat{\alpha}_{\perp\mu} \hat{\alpha}_{\perp}^\mu \right]
+ F_\sigma^2 \, \mbox{tr}
\left[ 
  \hat{\alpha}_{\parallel\mu} \hat{\alpha}_{\parallel}^\mu
\right]
+ {\cal L}_{\rm kin}(\rho_\mu) \ ,
\label{Lagrangian}
\end{equation}
where ${\cal L}_{\rm kin}(\rho_\mu)$ denotes the kinetic term of the
gauge boson $\rho_\mu$ of the HLS (vector meson) and
\begin{eqnarray}
\hat{\alpha}_{\stackrel{\perp}{\scriptscriptstyle\parallel}}^\mu 
&=&
\left( 
  D^\mu \xi_{\rm L} \cdot \xi_{\rm L}^\dag \mp
  D^\mu \xi_{\rm R} \cdot \xi_{\rm R}^\dag 
\right)
/ (2i) \ .
\end{eqnarray}
Two SU($N_f$)-matrix valued variables $\xi_{\rm L}$ and 
$\xi_{\rm R}$ transform as
$\xi_{\rm L,R}(x) \rightarrow \xi_{\rm L,R}^{\prime}(x) =
h(x) \xi_{\rm L,R}(x) g^{\dag}_{\rm L,R}$,
where $h(x) \in H_{\rm local}$ and 
$g_{\rm L,R} \in G_{\rm global}$.
These variables are parameterized as
$\xi_{\rm L,R} = e^{i\sigma/F_\sigma} e^{\mp i\pi/F_\pi}$,
where $\pi = \pi^a T_a$ denotes the Nambu-Goldstone (NG) 
boson associated with 
the spontaneous breaking of $G$ chiral symmetry and 
$\sigma = \sigma^a T_a$ 
denotes the NG boson absorbed into the HLS gauge boson.
$F_\pi$ and $F_\sigma$ are relevant decay constants, and
the parameter $a$ is defined as
$a \equiv F_\sigma^2/F_\pi^2$.
The covariant derivatives of $\xi_{\rm L,R}$ are defined by
$D_\mu \xi_{\rm L,R} =
\partial_\mu \xi_{\rm L,R} - i g \rho_\mu \xi_{\rm L,R}$,
where $g$ is the HLS gauge coupling.
In this paper we use $\pi$ for
the pseudoscalar NG bosons associated with the $G$ breaking
and $\rho$ for the HLS gauge bosons even for $N_f \neq 2$.

We adopt the background gauge field method
to obtain quantum corrections to the parameters.
This method was used in the chiral perturbation theory
(ChPT)~\cite{GL}, and
was applied to the HLS in Ref.~\cite{Tanabashi}.
We define two decay constants $F_\pi$ and $F_\sigma$ from 
two-point functions of the background fields 
$\hat{\alpha}_{\perp}^\mu$ and $\hat{\alpha}_{\parallel}^\mu$,
respectively.
The HLS gauge coupling $g$
is determined from two-point function of the
gauge field strength.
We note that the would-be NG boson $\sigma$ does contribute in the
$R_\xi$-like gauge fixing~\cite{HY} or the background gauge
fixing~\cite{Tanabashi}.

To handle the quantum effects properly
and relate parameters in the bare theory
defined at the cutoff scale with
those in the quantum theory at lower energy scale, 
we use the RGE's
in the Wilsonian sense which  
include the {\it quadratic divergences}
in addition to the logarithmic divergences.
As is usual, inclusion of only the logarithmic divergences is not
adequate to study the phase structure.
Actually, in the above  $CP^{N-1}$ model 
in $D$ ($2<D<4$) dimensions
it was essential to include
the power divergence which is the quadratic divergence in 4 dimensions.
As we shall show, the vector limit is still the fixed point
even if we include the quadratic divergences.

\begin{figure}[tbhp]
\begin{center}
\epsfxsize = 6.5cm
\ \epsfbox{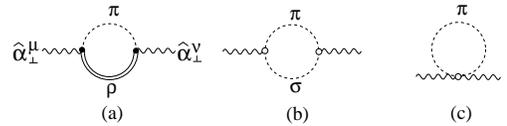}
\end{center}
\caption[]{Diagrams contributing to two-point function of
$\hat{\alpha}_{\perp}^\mu$.}
\label{fig:aa}
\end{figure}

\begin{figure}[tbhp]
\begin{center}
\epsfxsize = 7.5cm
\ \epsfbox{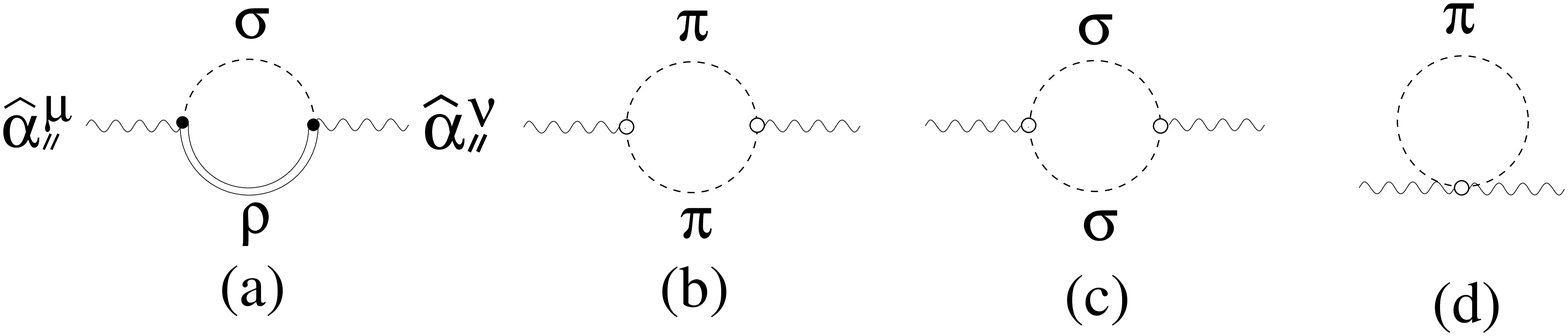}
\end{center}
\caption[]{Diagrams contributing to two-point function of
$\hat{\alpha}_{\parallel}^\mu$.}
\label{fig:vv}
\end{figure}

The RGE's for 
$g$ and $a$ above the $\rho$ mass scale
with only the logarithmic divergences are shown in
Ref.~\cite{HY}
where the parameters renormalized in the mass independent scheme were
studied and the vector limit was shown to be
realized in the
high energy limit.
Here we further include the quadratic divergences, since
we need RGE's in the Wilsonian sense
to study the phase structure.
Since a naive momentum cutoff violates the chiral symmetry,
we need a careful treatment of the quadratic divergences.
Thus we adopt the dimensional regularization
and identify the
quadratic divergences with the presence of poles
of ultraviolet origin at $n=2$~\cite{Veltman}.
We show the
diagrams contributing to two-point functions of
$\hat{\alpha}_{\perp}^\mu$ and
$\hat{\alpha}_{\parallel}^\mu$ in 
Figs.~\ref{fig:aa} and \ref{fig:vv}~\cite{HS}.
The resultant RGE's above the $\rho$ mass scale
are given by
\begin{eqnarray}
\mu \frac{d F_\pi^2}{d\mu} &=& 
C \left[ 3 a^2 g^2 F_\pi^2 + 2 (2-a) \mu^2 \right] \ ,
  \label{RGE for Fpi2}\\
\mu \frac{d g^2}{d\mu} &=& - C \frac{87 - a^2}{6} g^4 \ , 
  \label{RGE for g2}\\
\mu \frac{d a}{d\mu} &=& - C (a-1)
\left[ 3 a (a+1) g^2 - (3a-1) \frac{\mu^2}{F_\pi^2} \right] \ ,
  \label{RGE for a}
\end{eqnarray}
where 
$C = N_f/(2 (4\pi)^2)$ and 
$\mu$ is the renormalization scale.
The first term of Eq.~(\ref{RGE for Fpi2}) comes from the logarithmic
divergences of the diagrams in Figs.~\ref{fig:aa}(a) and (b), 
while the second
term comes from the quadratic divergences of the diagrams in
Figs.~\ref{fig:aa}(b) and (c).
The RGE for $a$ is obtained from those for $F_\pi$ and $F_\sigma$
through the definition $a \equiv F_\sigma^2/F_\pi^2$.
We note here that the above RGE's agree with 
those obtained in Ref.~\cite{HY}
when we neglect the quadratic divergences.
As is easily read from the RGE's~(\ref{RGE for g2}) and
(\ref{RGE for a}) 
the Georgi's vector limit~\cite{Georgi} 
$(g,a)=(0,1)$ is the fixed point.
The mass of $\rho$
is determined by the on-shell condition:
$m_\rho^2 = a(m_\rho) g^2(m_\rho) F_\pi^2(m_\rho)$.

Now, let us study how the quantum theory approaches 
to the chiral symmetric phase when $N_f$ becomes larger.
Here and henceforth
we write the dependence on $N_f$ 
as well as on the scale $\mu$ explicitly.
$F_\pi$ in the bare theory can be identified with that at the 
cutoff scale in the Wilsonian renormalization scheme.
This cutoff scale, say $\Lambda$, 
generally depends on $N_f$, so we express this by
$\Lambda_f \equiv \Lambda(N_f)$.
As we stated before,
the bare theory is written as if it were in the broken phase.
Then the parameter $F_\pi$ at the cutoff scale
does not vanish, and
it is natural to assume that $F_\pi(\Lambda_f)$ is of order of
$\Lambda_f$:
\begin{equation}
F_\pi(\Lambda_f;N_f) \sim \Lambda_f \ .
\end{equation}
Actually, the phase is determined by studying whether 
$F_\pi(0;N_f)$, which is the decay constant of NG bosons in the
quantum theory, vanishes or not.
The order parameter $F_\pi(0;N_f)$
will vanish due to the loop effects of $\rho$ and 
$\pi$~\cite{foot:1}.
This phenomena actually
occurs if the bare theory approaches to the vector
limit as $N_f$ becomes large.
Since the vector limit is the fixed point,
we may solve the RGE for $F_\pi$ Eq.~(\ref{RGE for Fpi2})
in the vector limit  with taking $(g,a)=(0,1)$.
This RGE tells us that
$F_\pi$ does not diverge, and hence we conclude that 
$m_\rho=0$ in the vector limit.
Thus the RGE~(\ref{RGE for Fpi2}) with $(g,a)=(0,1)$
relates the order parameter $F_\pi(0;N_f)$ with 
$F_\pi(\Lambda_f;N_f)$ as~\cite{foot:2}
\begin{equation}
\frac{F_\pi^2(m_\rho=0;N_f)}{\Lambda_f^2} = 
\frac{F_\pi^2\left(\Lambda_f;N_f\right)}{\Lambda_f^2}
- \frac{N_f}{2(4\pi)^2} \ .
\label{RGE for fpi2 at vector limit}
\end{equation}
Since $F_\pi(\Lambda_f;N_f) \sim \Lambda_f$,
right-hand side (RHS)
of Eq.~(\ref{RGE for fpi2 at vector limit})
will vanish for a large value of $N_f$.
The chiral symmetry in the quantum theory
is restored at a certain flavor $N_f^{\rm cr}$
when we take the vector limit in the bare theory:
\begin{equation}
F_\pi(0;N_f)/\Lambda_f
\mathop{\longrightarrow}_{N_f \rightarrow N_f^{\rm cr}} 0 \ .
\end{equation}
This is our main result.

Let us next calculate the critical flavor $N_f^{\rm cr}$.
Here we use the following ``physical'' inputs for $N_f=3$
with taking $a=1$~\cite{foot:3}:
$F_\pi(0;N_f=3) = 88$\,MeV determined in the chiral 
limit~\cite{GL};
$m_\rho(N_f=3) = 770$\,MeV;
$\Lambda(N_f=3) = 4 \pi F_\pi(0;N_f=3) \simeq 1.1$\,GeV 
from the naive dimensional analysis~\cite{MG}.
Below the $m_\rho$ scale, 
$\rho$ decouples 
and hence $F_\pi$ runs by the loop effect of $\pi$ alone.
The relevant Lagrangian with least derivatives is given by
the first term of Eq.~(\ref{Lagrangian}),
and the diagram contributing to $F_\pi^2$ is shown in
Fig.~\ref{fig:aa}(c).
The resultant RGE for $F_\pi$ is given by
$ (d F_\pi^2)/(d \mu^2) = 2 C$.
Then the order parameter 
$F_\pi(0;N_f)$ is related to $F_\pi(m_\rho(N_f);N_f)$ by
\begin{equation}
F_\pi^2(0;N_f) 
= 
F_\pi^2(m_\rho(N_f);N_f) - 
\frac{N_f}{(4\pi)^2} m_\rho^2(N_f) \ .
\label{sol fpi2 for chpt}
\end{equation}
The above input leads to 
$F_\pi(\Lambda_3;N_f=3) = 171$\,MeV and
$g(m_\rho;N_f=3) \simeq 5.6$, the latter of which is
consistent with the values of $g$ determined by assuming the
saturation of the ChPT parameter 
$L_9(m_\rho)$
by the vector mesons~\cite{EGLPR}; 
$g(m_\rho;N_f=3) = 6.0 \pm 0.4$.
For simplicity we assume
that $\Lambda_f$ and 
$F_\pi(\Lambda_f;N_f)/\Lambda_f$ do
not depend on $N_f$; 
$F_\pi^2(\Lambda_f;N_f)/\Lambda_f^2 \simeq0.024 $.
Then the critical flavor is determined
from Eq.~(\ref{RGE for fpi2 at vector limit})
as
\begin{equation}
N_f^{\rm cr}\simeq7.6 \ ,
\end{equation}
which is somewhat similar to the lattice calculation~\cite{IKKSY}.

\begin{figure}[htbp]
\begin{center}
\epsfxsize = 7.5cm
\ \epsfbox{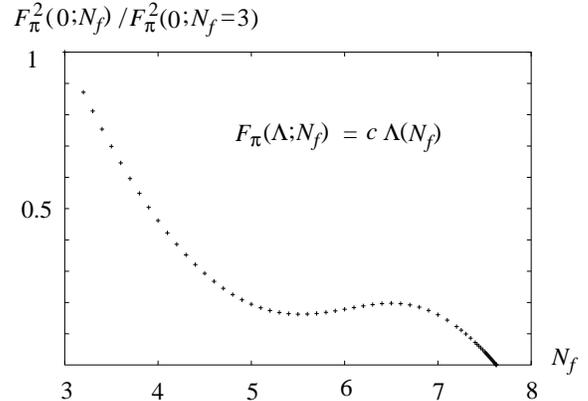}
\end{center}
\caption[]{$N_f$-dependence of $F_\pi^2(0;N_f)$,
normalized by $F_\pi^2(0;N_f=3)$.
The constant $c\simeq0.16$ is determined 
from the physical inputs for $N_f=3$
discussed in the text.}
\label{fig:3}
\end{figure}

To study how $F_\pi$ approaches to zero as $N_f$ is increased
we first need to determine
how the bare parameters
$g(\Lambda_f;N_f)$ and $a(\Lambda_f;N_f)$ approach to the values in 
the vector limit~\cite{foot:4}.
In the present analysis 
let us fix $a \equiv 1$
for simplicity~\cite{foot:3.2}.
We adopt the following behavior of the gauge coupling
approaching to zero:
\begin{equation}
g^2(\Lambda_f;N_f) 
= 
\bar{g}^2 \epsilon \ , \quad
\epsilon \equiv
1/N_f - 1/N_f^{\rm cr} \ ,
\label{initial value of g}
\end{equation}
where $\bar{g}$ is independent of $N_f$~\cite{foot:5}.
We present numerical calculation of the $N_f$-dependence of
$F_\pi(0;N_f)$ in Fig.~\ref{fig:3}.
This clearly shows that $F_\pi(0;N_f)$ 
smoothly goes to zero~\cite{foot:6}
at $N_f^{\rm cr} \simeq 7.6$.
Thus we conclude that 
{\it
the quantum theory provides the chiral restoration
when the bare theory approaches to the vector limit 
as $N_f \rightarrow N_f^{\rm cr}$}.

Several comments are in order:

In this paper we numerically studied the 
$N_f$-dependence of $F_\pi(0;N_f)$.
However, the RGE's are analytically solvable when we take $a=1$, and
the critical behaviors of $F_\pi(0;N_f)$ and $m_\rho(N_f)$ can be
studied analytically.
A careful analysis~\cite{HY:2} of the solutions of RGE's with the 
condition~(\ref{initial value of g}) leads to the fact that
when $N_f \rightarrow N_f^{\rm cr}$,
$g^2(m_\rho(N_f);N_f) \sim \epsilon$,
$F_\pi^2\left(m_\rho(N_f);N_f\right) /\Lambda_f^2
\sim \epsilon$, and hence
$m_\rho^2(N_f)/\Lambda_f^2 \sim \epsilon^2$.
This implies that
the second term of RHS of Eq.~(\ref{sol fpi2 for chpt})
approaches to zero faster than the first term.
Thus we obtain the critical behavior of the order parameter as
$F_\pi^2(0;N_f)/\Lambda_f^2 \sim \epsilon$~\cite{foot:7}.
This shows that the physical parameters $F_\pi(0)$ and $m_\rho$
approaches to zero in the power behavior,
which is originated from the fact that
we have used the one-loop perturbative RGE's.
If, on the other hand, we use some non-perturbative treatment,
we might obtain
an essential singularity scaling shown by an analysis of the SD
equation~\cite{ATW,MY}.

In the present analysis we took
$F_\pi(\Lambda_f)/\Lambda_f$ as well as $\Lambda_f$
as a quantity independent of $N_f$.
At the scale of $\Lambda_f$ we would like to match the HLS with QCD,
so that $N_f$-dependence of $\Lambda_f$ may be extracted from QCD.
However, as we can easily read from 
Eq.~(\ref{RGE for fpi2 at vector limit}),
imposing an $N_f$-dependence (increasing or decreasing)
of $\Lambda_f$, with $F_\pi(\Lambda_f)/\Lambda_f$ fixed, does not
change the critical flavor. 
On the other hand, if $F_\pi(\Lambda_f)/\Lambda_f$ depends on $N_f$,
the critical flavor will be slightly changed.
For example,
we can include the effect of the anomalous dimension obtained from QCD 
by using the Pagels-Stokar formula for $F_\pi(\Lambda_f)$ integrated
over the region $p^2 > \Lambda_f^2$.
This changes the resultant
value of $N_f^{\rm cr}$ to $N_f^{\rm cr} \simeq 6.5$.
A detailed
analysis will be shown in the forthcoming paper~\cite{HY:2}.

Axialvector mesons such as $a_1$ are heavier than the cutoff
scale, $\Lambda(N_f=3) \simeq 1.1$\,GeV, so that we did not include
them here.
On the other hand,
the recent analyses~\cite{HSS} show that there exist light 
scalar mesons.
Since the phase transition is expected to be
the conformal phase transition,
those masses as well as the baryon masses
will be small anyway~\cite{Chivukula}
when $N_f$ approaches to $N_f^{\rm cr}$.
Thus the critical flavor obtained in this paper $N_f^{\rm cr} \simeq
7.6$ might be changed by including those effects.

Although we can study the phase transition only from the broken phase
in this framework, it would be interesting to see whether the HLS
still makes sense even in the conformal window.

Our result may be applied to the dynamical electroweak symmetry
breaking such as technicolor.
As we discussed above, the $\rho$ (technirho) mass 
near the critical point becomes much smaller 
than $F_\pi(0;N_f)$
which is fixed to be the weak scale.
Thus we expect the 
light technirho as a signal
in the future collider experiments.


\begin{thebibliography}{99}
\bibitem{foot:1}
If we neglected the contributions from $\rho$, 
the relation between $F_\pi(\Lambda_f;N_f)$ and
$F_\pi(0;N_f)$
would be obtained by replacing $m_\rho(N_f)$ with $\Lambda_f$ in 
Eq.~(\ref{sol fpi2 for chpt}).
Then the chiral restoration would have occurred for $N_f \sim 4$.
However, the vector meson is in fact lighter than the cutoff 
scale already for $N_f=3$,
and hence we cannot neglect the contributions from $\rho$.
\bibitem{foot:2}
Here $\sigma$ does contribute 
even for $(g,a)=(0,1)$, since $\sigma$ is the real NG boson.
Thus there is an extra factor $1/2$ in the second term of 
Eq.~(\ref{RGE for fpi2 at vector limit}) compared with that of
Eq.~(\ref{sol fpi2 for chpt}).
\bibitem{foot:3}
In this paper the parameters are defined in the Wilsonian sense, and
we study the parameter $a$ only above
the $\rho$ mass scale $m_\rho$.
The vector meson dominance of the pion form factor, which leads to
$a=2$, was argued in the low energy region.
In the presence of the quadratic divergence, however,
the value of the parameter $a$ above the scale of $m_\rho$
differs from 2.
To give a quantitative feeling
let us naively extend the present 
RGE for $a$ below $m_\rho$ by including only
the $\pi$ effect shown in Fig.~\ref{fig:vv}(d).
Then we obtain $a(0;N_f=3) \simeq 1.8$ even
for $a(m_\rho;N_f=3)=1$.
\bibitem{foot:4}
Here we impose that the bare theory approaches to the vector limit at
$N_f^{\rm cr}$, although it would be possible that the vector limit
was already realized at $N_f < N_f^{\rm cr}$.
\bibitem{foot:3.2}
When we include a deviation of $a(m_\rho;N_f=3)$ from one,
the parameter $a$ runs by the RGE~(\ref{RGE for a}).
However, the value of the critical flavor 
as well as the qualitative structure 
changes only slightly.
(See Ref.~\cite{HY:2}.)
\bibitem{foot:5}
This behavior may be justified by the following large $N_f$ argument.
$g$ is the coupling of the three-point interaction of the vector
mesons.
Then large $N_c$ argument of QCD tells us that
$g^2$ behaves as $1/N_c$ in the large $N_c$ limit with fixed $N_f$.
On the other hand,
to make a large $N_f$ expansion in the HLS consistent the gauge
coupling $g^2$ falls as $1/N_f$.
However, the HLS is actually related to 
QCD, so that the large $N_f$ limit should be taken with
$N_c/N_f$ finite.
This situation can be seen by rewriting the RHS of
Eq.~(\ref{initial value of g}) into
$\left(\bar{g}^2/N_c\right) 
\left( N_c/N_f - N_c/N_f^{\rm cr} \right)$.
\bibitem{foot:6}
There is a small dip around $N_f = 5$.
As we stated in \cite{foot:1}, below the $m_\rho$ scale
$F_\pi$ decreases rapidly as $\mu$ decreases with fixed $N_f$
by the $\pi$ loop effect.
This effect becomes larger as $N_f$ increases.
Thus the chiral symmetry might have been restored by the effect.
However, for $N_f \gtrsim 5$
the $\rho$ mass becomes small rapidly enough to protect such a
situation to happen.
\bibitem{foot:7}
When we change the power of $\epsilon$ in 
Eq.~(\ref{initial value of g})
from one to $q$ ($q>1$),
$m_\rho(N_f)$ approaches to zero as
$m_\rho^2(N_f)/\Lambda_f^2 \sim \epsilon^{1+q}$, while
the critical behavior of $F_\pi(0;N_f)$ is not changed:
$F_\pi^2(0;N_f)/\Lambda_f^2 \sim \epsilon$.
(See Ref.~\cite{HY:2}.)
\end{thebibliography}
\end{document}